\documentstyle[12pt,aaspp4,rotating]{article}
\def\msun{{\,M_\odot}}

\def\simlt{\lower.5ex\hbox{$\; \buildrel < \over \sim \;$}}
\def\simgt{\lower.5ex\hbox{$\; \buildrel > \over \sim \;$}}

\def\cm{{\rm\,cm}}

\def\gcm3{{\rm\,g\,cm^{-3}}}
\def\ncm3{{\rm\,cm^{-3}}}

\def\>{$>$}
\def\<{$<$}

\def\mnras#1#2#3{#1, {\it MNRAS}, {\bf#2}, #3.}

\lefthead{Melia et al.}
\righthead{}

\received{2002 September 11}
\begin{document}
\centerline{Submitted to the Editor of the Astrophysical Journal}
\vskip 0.5in
\title{\bf Shadowing of the Nascent Jet in NGC~4261 \\
by a Line-Emitting Supersonic Accretion Disk}

\author{Siming Liu}
\affil{Department of Physics, The University of Arizona, Tucson, AZ 85721}

\author{Michael J. Fromerth\altaffilmark{1}}
\affil{Department of Physics, The University of Arizona, Tucson, AZ 85721}

\and

\author{Fulvio Melia\altaffilmark{2}}
\affil{Department of Physics and Steward Observatory, The University of Arizona, Tucson,
AZ 85721}
\altaffiltext{1}{NSF Graduate Fellow.}
\altaffiltext{2}{Sir Thomas Lyle Fellow and Miegunyah Fellow.}

\begin{abstract} 

NGC~4261 (3C~270) is a low-luminosity radio galaxy with two symmetric kiloparsec-scale
jets. Earlier {\it Hubble Space Telescope} observations indicated the presence of a
hundred-parsec scale disk of cool dust and gas surrounding a central, supermassive ($\sim
4.9\times 10^8\msun$) black hole.  The recent detection of free-free radio absorption by
a small, geometrically-thin disk, combined with earlier studies of the disk's large scale
properties, provide the strictest constraints to date on the nature of the accretion
process in this system.  We show here that a supersonic disk, illuminated by the active
galactic nucleus (AGN), can not only account for the observed radio shadowing, but can
also produce the optical broad lines emitted from this region.  
At large radii, the gas is optically-thin because the ram pressure due to turbulence 
is much larger than the thermal pressure of the gas. 
At smaller radii, but beyond a critical radius $r_c$, line cooling dominates over 
gravitational dissipation and the gas is effectively cooled down to temperatures below 
$10^4$ K.  
Within $r_c$, however, heating due to the release of gravitational energy
overwhelms line cooling and the gas falls onto the unstable portion of the cooling curve.  
Because cooling is quite inefficient under these conditions, the plasma is heated very
quickly to a temperature close to its virial value as it falls toward the central engine.
Thermal pressure of the gas dominates the turbulent ram pressure at a radius $\sim (2/3)\
r_c$, below which the flow probably becomes advection dominated.  The disk is
optically-thin to UV and X-ray radiation within $r_c$, so the ionizing radiation from the
AGN is preferentially absorbed near $r_c$, affecting the disk structure significantly.  
To include the ensuing photoionization effect, we have used the algorithm Cloudy with
additional heating introduced by gravitational dissipation to calculate the temperature
profile and line emission from the disk in a self-consistent manner.  The results of our
model calculation are consistent with current multiwavelength observations of the disk in
this source.

\end{abstract}


\section{Introduction}

The development of supermassive black hole accretion theories has always been limited by
observations (Pringle 1981).  This has been changing quickly during the past few years, with
increasingly higher resolution observations being made over a wide frequency range.  For
example, the {\it Chandra X-ray Observatory} ({\it Chandra}) and the {\it Hubble Space
Telescope} ({\it HST}) both have sub-arcsec scale resolutions, while the {\it Very Large
Baseline Interferometer} ({\it VLBI}) can resolve structures on sub-milliarcsec scale.  By
combining multiwavelength observations from these different instruments, we can obtain very
strict constraints on the nature of accretion around supermassive black holes.

While the innermost structures of AGNs remain unresolved (however, cf.~Melia \& Falcke 2001
regarding the Galactic center), the current data are good enough to constrain the properties
of the large-scale accretion disks that are presumably feeding the central engines.  A
better understanding of these large-scale disks is not only tenable, but will also provide
significant implications on the nature of the AGNs.  Most of the relevant theoretical work
to date is based on the standard $\alpha$-disk prescription (Shakura \& Sunyaev 1973), in
which the accretion disk at large radii is supported by the thermal pressure of the gas
(see, e.g., Hubeny 1990; Hur\'{e} et al.~1994; Collin \& Hur\'{e} 1999).  This may not be
the entire story, however, as some observations suggest that turbulence is prominent in the
disks of many radio galaxies (Ferrarese, Ford \& Jaffe 1996; Ferrarese \& Ford 1999; van
Langevelde et al.~2000).  In addition, the inner disk is irradiated (Maloney, Hollenbach \&
Tielens 1996), which should affect, e.g., its physical properties and line emission
characteristics.

At a distance of only $30$~Mpc, the FR~I radio galaxy NGC~4261 provides us an exceptional
opportunity to investigate these new effects. Harboring a $4.9\times 10^8\msun$ black hole
at its center (Ferrarese et al.~1996), NGC~4261 is well known for its kiloparsec-scale jets
(Birkinshaw \& Davies~1985) and its hundred parsec-scale dusty disk imaged by {\it HST}
(Jaffe et al.~1993). This disk possesses an atomic surface density of $\sim 5\times 10^{20}$
cm$^{-2}$, as inferred from its optical depth assuming a Galactic dust/gas ratio. On a
smaller scale, a thin circumnuclear disk of radius $\sim 6$~pc was revealed by radio line
absorption studies (Jaffe \& McNamara 1994; van Langevelde et al.~2000). This portion of the
disk has an \ion{H}{1} column density of $\sim 2.5\times 10^{19}\, T_{\rm sp}$~cm$^{-2}$,
where $T_{\rm sp}\simeq 100$ K is the spin temperature of the gas. At an even smaller scale,
free-free absorption of radio emission from the core by a geometrically-thin disk of radius
$\sim 0.3$~pc was detected (Jones et al. 1997, 2000, 2001; Piner, Jones \& Wehrle 2001).
This recent finding adds significant content to the accretion picture because this radius
corresponds to only $6400\,r_S$, where the Schwarzschild radius $r_S\equiv 2GM/c^2\simeq\
1.45\times 10^{14}$~cm for the central black hole in NGC~4261.

Although these observations of the accretion disk extend to sub-parsec scales, a fully
self-consistent accretion model does not yet exist, particularly for the small free-free
absorbing disk (Sudou et al.~2000). Jones et al.~(2000) did have success in attributing the
observed free-free absorption to a standard $\alpha$-disk. Unfortunately, to match
the observations, the model requires a dissipative stress that is much greater than
that provided by subsonic turbulence. A viable alternative for (or addition to) this 
stress must be found in order for the model to be fully self-consistent.

We suggest here that supersonic turbulence is the responsible agent. Observations of
NGC~4261 provide direct evidence that turbulence is prominent in the disk from $100$~pc to
about $6$~pc, with the inner boundary set by observational limitations. At 100~pc, the width
of narrow lines from the disk suggest a turbulent velocity of $70$~km/s (Jaffe et al.~1996),
while at 6~pc, the turbulent velocity estimated from the \ion{H}{1} absorption profile is
about $130$~km/s (Jaffe \& McNamara 1994; van Langevelde et al.~2000). Strong turbulence in
the accretion disk is understandable given that the disk is embedded within the core of the
galaxy, where it is subject to disruption by strong stellar winds.  The dissipation of
turbulence is regulated by the cascade in energy from large scales to small scales, so the
total energy associated with the turbulent motion decays roughly as a power-law in time
(Lazarian 2003), even in the case of supersonic turbulence (Smith, Mac Low \& Zuev 2000a).  
This slow turbulence decay suggests that the supersonic structure might persist even
to smaller radii.

In such a supersonic disk, the accreting gas is optically-thin and gravitational energy
dissipation will overwhelm line cooling at a critical radius $r_c$, within which the gas is
effectively heated to a temperature close to its virial value (Pringle 1981).  Thermal
pressure of the gas will then dominate the turbulent pressure at $r \simeq (2/3)r_c$, so the
supersonic flow appears to terminate at this radius and the gas may make a transition into
an advection-dominated accretion flow thereafter (Narayan \& Yi 1994).  On larger scales,
the supersonic structure can be inferred directly from observations.  It is interesting to
note that with a stress parameter $\alpha \simeq 1$, the value of $r_c$ is roughly $0.3$ pc,
which corresponds to the inferred size of the free-free absorbing region. Thus the observed
shadowing of the nascent radio jet may be produced by the ionized gas near $r_c$.

We further note that newly released {\it Chandra} data also hint that the nucleus 
of NGC~4261 may be
weakly-active (Peter, Zezas \& Fabbiano 2001), consistent with earlier {\it ASCA} and {\it
ROSAT} observations (Matsumoto et al.~2001; Worrall \& Birkinshaw 1994).  The corresponding
ionization parameter at $0.3$~pc (assuming that the radiation is isotropic) is $\Xi \approx
5$, where $\Xi\ \equiv\ F_{\rm ion}\, /\, n_{\rm tot} k T c$, $F_{\rm ion}$ is the ionizing
flux between 1 and $10^3$~Ryd, and $n_{\rm tot}$ is the total number density including all
species.  Note that $\Xi$ roughly corresponds to the ionizing-photon-to-gas pressure ratio.
A value of $\Xi \gtrsim 1$ indicates that the gas may be effectively photoionized.  This
photoionization will significantly affect the profiles of the lines emitted by the
supersonic disk. High-resolution {\it HST} imaging of the central ionized gas has detected
broad emission lines from a region smaller than 7 pc in radius (Ferrarese et al.~1996).  In
this paper, we show that a supersonic disk, irradiated by the AGN, can not only account for
the observed radio absorption feature, but also reproduce these broad emission lines.  We
also show that its large-scale extension is consistent with the radio line absorption
observations.

The physical conditions present in the broad line regions (BLRs) of this class of radio
galaxy are likely to be very different than those present in the more-luminous Seyfert
galaxies and quasars. In both scenarios, the line emission is produced by gas having
temperature on the order of a few times $10^4$~K, with line broadening caused by bulk
motions of the emitting plasma. However, while the BLR in Seyferts and quasars is primarily
photoionized, our model predicts that heating due to gravitational dissipation plays a
significant role in the low-luminosity radio galaxies.

Our modeling also indicates a very different BLR geometry for these galaxies, with the line
emission coming directly from the turbulent accretion disk. While accretion disks may play
an important role in some BLR models of Seyferts and quasars, the line emission is not
believed to originate in the disk itself, instead being associated with, e.g.,
magnetohydrodynamic flux tubes (Emmering, Blandford, and Shlosman 1992) or radiation- and
gas-pressure driven winds rising from the surface of the disk (Murray et al.~1995). The
accretion disk does not even play a direct role in many successful BLR models for these
objects, e.g., the bloated star (Alexander \& Netzer 1994) or accretion-shock cooling models
(Fromerth \& Melia 2001).

It is also noted that the density of the BLR gas in these galaxies is much lower than in
traditional AGNs. Because broad forbidden emission lines are absent in Seyfert and quasar
spectra, we know that the density of their line-emitting gas is $n \gtrsim 10^7$~cm$^{-3}$.
However, the spectra of some low-luminosity radio galaxies contain strong broad forbidden
emission lines (e.g., [\ion{N}{2}], [\ion{S}{2}]), indicating gas densities $n \lesssim
10^5$~cm$^{-3}$ (see, e.g., Ferrarese \& Ford 1999; Barth et al. 1999). The lower density 
BLR gas may be directly tied to their lower mass accretion rates and resulting luminosities.

We first discuss the general properties of a supersonic disk in \S\ \ref{sec:turbdom}. Its
consistency with the multi-wavelength observations of NGC~4261 is discussed in \S\
\ref{estimate}. Detailed modeling of the line emission from this supersonic disk, irradiated
by its AGN, is given in \S\ \ref{lineemission}. In \S\ \ref{discussion}, we discuss the
implication of this work on the nature of AGNs in radio galaxies.

\section{A Supersonic Accretion Disk Model}
\label{sec:turbdom}

Because it is thought that the large-scale dusty disk feeds the inner accretion of the AGN
(Jaffe et al.~1996), the detection of a prominent gap in the radio core of NGC~4261 prompted
Jones et al.~(2000, 2001) to introduce a free-free absorbing disk to account for the
observed features.  To produce the radio absorption, the inner disk must have an emission
measure of $n_e^2 \, l \simeq 3\times 10^8\ T_4^{3/2}$~pc~cm$^{-6}$, where $n_e$ is the
electron density, $l$ is the path length through the absorbing gas, and $T_4$ is the
electron temperature in units of $10^4$~K.  If dominated by thermal gas pressure, however,
the optically thin $\alpha$-disk proposed by these authors cannot account for this radio
shadowing in a self-consistent manner. In particular, we show below that the implied
stress parameter $\alpha$ is too large for the flow to be subsonic.

For a gas-pressure-dominated accretion disk, angular momentum conservation requires 
that $\nu \simeq -(2/3)\,v_r\,r$ at large radii, where $\nu = \alpha\,H(r)\,c_s(r)$
is the kinetic viscosity, $\alpha$ is the stress parameter
(assumed to be radius-independent and less than one), $v_r$ is the radial velocity 
of the flow, $H(r)$ is the scale height of the disk, and $c_s(r)$ is the local sound  
speed. From the expressions given in Jones et al.~(2000), one can obtain:
\begin{equation}
\alpha \ \simeq \ - \frac{2}{3} \frac{v_r \ r}{H\ c_s} \ = \ -\frac{2}{3} \
\frac{v_r}{v_k} \left({r\over H}\right)^2 \ =\  2.4\times 10^3 \, M_8^{1/4} \,
\dot{M}_{-3}^{1/2} \, r_{18}^{-3/4} \ ,
\end{equation}
where $v_k=(GM/r)^{1/2}$ is the Keplerian velocity, $M_8$ is the black hole
mass in units of $10^8\msun$, $\dot{M}_{-3}=\dot{M}/(10^{-3}\msun / {\rm
yr})$ gives the accretion rate, and $r_{18}\sim 1$ is the distance
of the radio absorbing gas from the black hole in units of $10^{18}$ cm. In
fact, it can be shown that the disk cannot be supported by thermal
pressure of the gas alone since the corresponding accretion rate is:
\begin{eqnarray}
\dot{M}\
& \simeq\ & 7.4\times 10^{19}\ \alpha\
\left(n_o\over 2.0\times 10^5~\cm^{-3}\right)
\left(T_o\over 10^4\ {\rm K}\right)^{3/2}
\nonumber \\
&&\times
\left(M\over 4.9\times 10^8\ \msun\right)^{-1}
\left(r_o\over 10^{18}\ \cm\right)^3 {\rm g\ s^{-1}} \ ,
\end{eqnarray}
where quantities with subscript $o$ denote the corresponding values inferred from the radio 
absorption features, and we have approximated the line-of-sight path length through the disk 
with $2H/\sin{(26^\circ)}$ (Jones et al. 2001), where $H = (R_gT r^3/\mu GM)^{1/2}$, $R_g$ 
is the gas constant and $\mu = 0.5$ is the molecular weight. 
Such an accretion rate is so small that the implied emission efficiency for converting 
accreted rest-mass energy into X-ray emission from the nucleus of NGC~4261 is about 
$3.0/\alpha$ (Worrall \& Birkinshaw 1994; Matsumoto et al.~2001), 
which is unacceptable for any $\alpha\le 1$ in a thermal pressure supported disk.  
Sudou et al.~(2000) further discussed an optically-thick standard disk and an optically-thin 
disk cooled by free-free emission. Neither of these alternatives can account for the 
observed shadowing.

Meanwhile, several observations indicate that turbulence dominates the gas motion at large 
scales (Jaffe et al.~1996; van Langevelde et al.~2000), where supersonic turbulence may be 
produced by stellar winds in the core of the galaxy (Jaffe et al. 1996). 
As the gas falls in toward the black hole, however, the strength of the residual turbulence 
in the disk is determined by balancing shock dissipation with the turbulence enhancement due 
to gravitational energy dissipation. In a fully developed turbulent flow, energy is mostly 
associated with large-scale gas motion. The decay rate of the total turbulent energy is then 
determined by the rate at which energy cascades from large scales to small scales. 
The turbulent energy density per unit mass, $\varepsilon$, is therefore given by 
\begin{equation}
\dot{\varepsilon} = \varepsilon/\tau\,,
\end{equation}
where the dot denotes a derivative with respective to time and $\tau = L/\varepsilon^{1/2}$ 
gives the characteristic time scale for transporting energy from large to small scales. 
The characteristic length scale, $L$, of the system depends on the turbulence generation 
mechanism. Assuming $L$ is independent of time, we have $\varepsilon \propto t^{-2}$. 
In their numerical simulations of the decay of supersonic turbulence, Smith et al.~(2000a) 
found that the total energy of the turbulence decays roughly with a power law in time, with 
an index $-1.5$.

We point out that as the large-scale supersonic gas flows inward, the dissipative stress is 
provided primarily by its turbulent motion. So the dissipated gravitational energy will go 
first into producing turbulence. This turbulence generation mechanism can, in principle, 
keep the disk supersonic even on smaller scales (Smith, Mac Low \& Heitsch 2000b). A 
complete model addressing the evolution of this supersonic accretion flow, however, would 
require a detailed numerical simulation beyond the scope of this paper.

In the following discussion, we will assume that the energy associated with the turbulent 
motion is in sub-equipartition with, and proportional to, the gravitational energy of the 
gas. We therefore have
\begin{equation}
v_t^2\ =\ f\, v_k^2 \ ,
\end{equation}
where $f$ is a ``turbulence parameter'', which can be fixed by the observed disk properties 
at large radii, and $v_t$ is the turbulent velocity in the flow. The supersonic disk model 
(in which the stress is dominated by turbulence) then gives:
\begin{eqnarray}
\nu &=& \alpha\, H\, v_t\ , \\
H &=& f^{1/2}\, r\ , \label{height} \\
v_r &=& -1.5\ {\nu\over r}  \nonumber \\
 & = & -1.5\ \alpha\, f\, v_k\ , \\
\dot{M} &=& 6\pi\ \alpha\, f^{3/2}\, r^2\, v_k\, n\, m_p \ .
\label{numd}  
\end{eqnarray}
In Eq.~(\ref{numd}), $n(r)$ is the baryon number density averaged in the vertical
direction of the accretion disk and $m_p$ is the proton mass.  Because the {\rm H}
{\small I} absorption feature indicates that $v_t \simeq 0.21\ v_k$ at $r \simeq 6$~pc 
from the nucleus, we obtain $f \simeq 0.045$.  The corresponding disk opening angle 
is $2\,\tan^{-1}f^{1/2}\ \simeq\ 24^\circ$.  Then from Equations (\ref{height}) 
and (\ref{numd}), we have
\begin{eqnarray}
H&=&0.21\ r\ , \\
n&=&n_o\left({r_o\over r}\right)^{3/2}\ . \label{eq:density}
\end{eqnarray}
It is interesting to note that the viscous time scale in such a disk ($t_{vis}=r/v_r$ by 
definition) is longer than the corresponding dynamical time scale ($t_d = 2\pi r/v_k$) by 
a factor of $2.4/\alpha$. Adopting the power-law decay rate from the numerical 
simulations (Smith et al. 2000a), one can show that the ratio of the turbulent energy to 
the dissipated gravitational energy is about $0.27 \alpha^{3/2}$. This is consistent with 
$f\approx 0.045$ for $\alpha$ less than one.

One should emphasize that only a small fraction ($\sim 4.5\%$) of the dissipated 
gravitational energy is left to drive the turbulent motion.  Most of the dissipated 
gravitational energy is thermalized via shocks and effectively radiated away. Since we 
are mostly interested in regions where the gas is partially ionized, this cooling is 
dominated by line emission. We will see that there is a critical radius $r_c$ in the disk, 
where the gravitational energy dissipation rate equals the line cooling rate.
The gravitational energy dissipation rate is given by
\begin{equation}
\Gamma = \frac{3}{8 \pi} \frac{G\, M\, \dot{M}}{H\, r^3} \ .
\label{heat}
\end{equation}
From Equation (\ref{numd}), we have
\begin{eqnarray}
\dot{M}\
& = & \ 2.3\times 10^{24}\ \alpha\,
\left({f\over 0.045}\right)^{3/2}
\left({n_o\over 3\times 10^4\ \cm^{-3}}\right) \nonumber \\
&   & \times
\left({r_o\over 10^{18}\ \cm}\right)^{3/2}
\left({M\over 4.9\times 10^8\ \msun}\right)^{1/2}
{\rm g\ s^{-1}}\ . \label{mdot}
\end{eqnarray}
Using this equation to eliminate $\dot{M}$ in Equation (\ref{heat}), we get
\begin{eqnarray}
\Gamma &=& 8.5\times 10^{-14}\ \alpha\, \left[\
\left({f\over 0.045}\right)
\left({n_o\over 3\times 10^4\ \cm^{-3}}\right)
\left({r_o\over 10^{18}\ \cm}\right)^{3/2} \right. \nonumber \\
&& \times \left. \left({r\over 10^{18}\ \cm}\right)^{-4}
\left({M\over 4.9\times 10^8\ \msun}\right)^{3/2}\  \right]\
{\rm ergs\ cm^{-3}\ s^{-1}}\ .
\label{eq:addheat}
\end{eqnarray}
The line cooling rate of an optically-thin gas is given by
\begin{eqnarray}
\Lambda\ \equiv\ n^2 \lambda&=&9.0\times 10^{-14}\ \lambda_{-22}\
\left[\ \left({n_o\over 3\times 10^4\ \cm^{-3}}\right)^2
 \right. \nonumber \\
&& \times \left. \left({r_o\over 10^{18}\ \cm}\right)^{3} \left({r\over 10^{18}\
\cm}\right)^{-3}\ \right]
\ {\rm ergs\ cm^{-3}\ s^{-1}},
\end{eqnarray}
where $\lambda$ is the cooling function (Krolik 1999) and $\lambda_{-22} \equiv
\lambda / (10^{-22}\ {\rm ergs\ cm^3\ s^{-1}})$.

In terms of the critical radius $r_c$, we have
\begin{eqnarray}
\alpha
\left({f\over 0.045}\right)
\left({M\over 4.9\times 10^8\ \msun}\right)^{3/2}
& = & 1.06\ \lambda_{-22}
\left({n_o\over 3\times 10^4\ \cm^{-3}}\right) \nonumber \\
&& \times
\left({r_o\over 10^{18}\ \cm}\right)^{3/2}
\left({r_c\over 10^{18}\ \cm}\right)\ . \label{critical}
\end{eqnarray}
To account for the radio shadowing, the electron number density should be $\sim
3\times 10^4$ cm$^{-3}$ at $\sim 0.3$ pc, where we have assumed an inclination angle
$i = 64^\circ$ for the disk (Jones et al.~2001).  Thus, for a stress parameter $\alpha
\simeq 1$, Equation (\ref{critical}) shows that the critical point is located within the
radio shadowing region.  It is natural to expect that the observed radio absorption 
should be produced by the fully ionized gas near $r_c$, and we will see below that
ionizing photons from the central engine play a crucial role in determining the
ionization state of the gas at this radius. 

{\it HST} observations indicate that the gas closest to the nucleus may be ionized by 
radiation from the central engine (Ferrarese et al.~1996). Further evidence for this can be 
shown by estimating the ionization parameter at $r_c$. Recent high-resolution {\it Chandra} 
observations have uncovered a high-energy source at the nucleus of NGC~4261 with a hard band 
X-ray luminosity of $L \approx 7 \times 10^{40}$~ergs/s (Peter et al.~2001), confirming 
earlier {\it ASCA} results (Matsumoto et al.~2001). Meanwhile, the soft X-ray luminosity is 
$\sim 10^{41}$ ergs/s (Worrall \& Birkinshaw 1994). So, assuming that the radiation is 
isotropic and that the gas temperature is $\sim 10^4$~K, we can estimate the value of the 
ionization parameter to be $\Xi \simeq 5$ for the radio-absorbing gas at $\sim
10^{18}$~cm (Krolik 1999). The radio-absorbing gas is therefore effectively photoionized.

Although the radiation pressure dominates the thermal pressure of the gas in the
radio absorbing region, the turbulent pressure there is $0.5\,n_o\,m_p\,v_t^2\,
\simeq\,7.4\times 10^{-5}$~ergs~cm$^{-3}$, which is much larger than the radiation
pressure.  So the disk structure, including the scale height and number
density, is still determined by the turbulent motion.  However, the gas temperature
profile depends on the interaction between the disk and the radiation field.  In the
following sections, we will incorporate this photoionization effect using the
algorithm Cloudy (Ferland 1996; Cloudy 96-$\beta$3) and show that the observed broad-line 
spectrum may also be fitted with this disk model in a self-consistent fashion.

\section{Line Emission from the Irradiated Supersonic Disk}
\label{estimate}

To model line emission from the irradiated disk, we first need to know the intensity
and spectral energy distribution (SED) of the ionizing radiation from the central
source.  These are highly uncertain due to a combination of factors. First, there 
are uncertainties associated with intrinsic absorption and galactic contamination 
which limit the precision to which we can quantify the spectrum in the optical and 
X-ray bands. Second, the spatial distribution of photons is unknown --- it is 
unclear whether the central engine emits isotropically; the disk at $r_c$ may 
``see'' a different spectrum from what we observe. Third, and most importantly, we 
do not have observations (or even good estimates) of the continuum at UV energies. 
Generally, the shape and intensity of the UV continuum most strongly determine the 
properties of a photoionized gas.

The correlations between radio and optical spectra have suggested a non-thermal
nuclear source for FR I radio galaxies (Chiaberge, Capetti \& Celotti 1999).  
Further evidence for a power-law nuclear source in NGC~4261 comes from {\it ROSAT} and {\it 
ASCA} observations in the X-ray band (Worrall \& Birkinshaw 1994; Matsumoto et
al.~2001), which are confirmed by recent high spatial resolution {\it Chandra}
observations (Rinn \& Sambruna 2001;  Peter et al.~2001).  In the following
discussion, we will adopt a power-law spectrum in the UV-range to bridge the gap
between the optical and X-ray data. The continuum is assumed to be emitted
isotropically from a central, point-like source, and the intensity of the ionizing
luminosity remains a free parameter in our model to reflect our uncertainties.

We are mostly interested in the region at $\sim 0.3$ pc.  However, to determine the
spectrum of the incident radiation, we also need to know the gas distribution in the
inner region since radiation from the core may be reprocessed by intermediate gases.
Fortunately, according to the supersonic disk model, the disk will be heated up very
quickly within $r_c$, which makes the gas there optically-thin to UV and X-ray
emission.  One can show this by solving the energy conservation equation in the
transition region between $r_c$ and the radius $r_t$, where the thermal pressure of 
the gas equals the turbulent pressure and the supersonic accretion flow
effectively terminates.  

For a supersonic accretion disk, we have the energy conservation equation:
\begin{equation}
{d\over dr}\left[ 1.5\, (1-f)\, \dot{M}\, v_k^2\ -\ \frac{5\, \dot{M}\, k_B\,
T}{m_p} \right] \ =\ -4\pi\ r\, H\, \Lambda \ .
\end{equation}
The first term on the left-hand side corresponds to the gravitational energy
dissipation and turbulent advection, while the second term is associated with
the thermal energy of the gas.  The right-hand side gives the cooling. The
quantity $\Lambda/n^2 \equiv \lambda$ is about $10^{-22}$ ergs cm$^3$ s$^{-1}$ near the 
peak (at $\sim
10^4$~K) of the cooling function (Krolik 1999). Assuming that $\lambda$
does not depend on radius, we have the following temperature profile in the transition
zone:
\begin{eqnarray}
T(r) &=& T_c\ +\ {0.3\ (1-f)\ G\, M\over r_c\ R_g}\ \left[\ {r_c\over r}\ -1\  +\ 
\ln
(r/r_c)\ \right]  \nonumber \\
&\simeq& T_c\ +\ {0.15\ (1-f)\ G\, M\over r_c\ R_g}\ \left[{r_c-r\over r}\right]^2
\nonumber \\
& = & T_c \ +\ 1.2\times 10^8\ (1-f)\,
\left({r_c-r\over r}\right)^2 \nonumber \\
& & \left({r_c\over 10^{18}\ \cm}\right)^{-1}
\left({M\over 4.9\times 10^8\ \msun}\right)\ {\rm K}\ . \label{temp}
\end{eqnarray}
In the second expression above, we have kept only the term proportional to 
$[(r_c-r)/r]^2$ and have neglected the high order terms in $(r_c-r)/r$.
Setting $2\ k_B\, T(r_t) = f\,m_p\,v_k^2$, we obtain the termination radius $r_t$ of
the supersonic flow:
\begin{equation}
r_t\  =\  \frac{r_c}{1\ +\ a/2\ +\ (a+a^2/4)^{1/2}} \ ,
\end{equation}
where
\begin{equation}
a\ =\ {f\over 0.3\ (1-f)}\ \simeq\ 0.15\, \left({f\over 0.045}\right).
\end{equation}
So $r_t\simeq (2/3)r_c$, and $T(r_t)\simeq 3\times 10^7$ K. Then the error
introduced by neglecting the high order terms in Equation (\ref{temp}) is
$\sim 30\%$. In general, the termination radius $r_t$ is determined by
solving the equation
\begin{equation}
\ln{\left(r_t\over r_c\right)}\ =\ 1 + {(8f\, -\, 3)\, r_c\over 3\, (1-f)\, r_t} \ .
\end{equation}
So $f$ must be smaller than 3/8 to make the thermal pressure of the gas dominate at
small radii.  A larger value of $f$ will cause the turbulent advection to cancel 
most of the gravitational energy dissipation and keep the disk supersonic.  In
addition, a more accurate expression for $\lambda$ in the transition zone will make
the gas heated even faster because it is on the unstable portion of the cooling 
curve (Liu, Fromerth \& Melia 2002). Within $r_t$, the accretion pattern may transfer 
into an advection dominated flow (Narayan \& Yi 1994). 

Within $r_c$, the disk is optically-thin to UV and X-ray radiation,
and because of the sharp increase in the gas temperature in this region, free-free
absorption is also negligible.  Due to the effects of photoionization by the AGN, 
however, the gas will be fully ionized well beyond $r_c$.  To incorporate this
effect self-consistently, we will need to use the algorithm Cloudy
with additional heating introduced by gravitational dissipation to calculate the
temperature and the ionization state of the gas everywhere in the disk.

Before calculating the line emission from the supersonic disk numerically, it may be
illustrative to estimate the physical conditions of the broad-line region. Observations
with {\it HST} have identified an ionized gas concentration in a spatially-resolved region 
with a FWHM of $17$ pc.  The broad emission lines are associated with the central
$0\farcs1\sim 14\ {\rm pc}$ aperture position (Farraese et al.~1996).  For the
H$\alpha$ line, the observed flux is $3.11\times 10^{-15}$~ergs~cm$^{-2}$~s$^{-1}$
with a FWHM of 2500~km/s.  From the electron number density inferred above and the
H$\alpha$ volume emissivity of $\sim 2\times 10^{-25}$~ergs~cm$^{3}$~s$^{-1}$
(Osterbrock 1974), we can estimate the size of the emission region.  In our model,
most of the lines should be produced beyond $r_o$. The radius $r$ of this region 
should therefore satisfy the condition 
\begin{equation}
4.8\times 10^{38}
\left({f\over 0.045}\right)^{-1/2}
\left({n_o\over 3\times 10^4\cm^{-3}}\right)^2
\left({r_o\over 10^{18}\cm}\right)^{3}
\ln{(r/r_o)}= 
3.3\times 10^{38}\;, \label{line}
\end{equation}  
implying that $r\approx 0.6$ pc. It is interesting to note that the Keplerian velocity at 
such a radius is 1900~km/s, which is consistent with the FWHM of the H$\alpha$ line if one 
takes into account the turbulent motion in the flow and the inclination of the disk.  
Because the contributions to H$\alpha$ from the various rings at different
radii are comparable to each other between $\sim 0.3$~pc and $\sim 0.6$~pc, 
the double peaks that are usually produced by Keplerian motion are here filled 
in, consistent with the observed spectrum.  So a supersonic accretion disk with 
the properties described above appears to account for the broad emission lines 
rather naturally.

\section{Modeling the Broad Emission-Line Spectrum}
\label{lineemission}

Several observational constraints already limit the parameter space available
for fitting the broad emission lines. Images from {\it HST} show that the large-scale
dusty disk has an inclination angle of $64^\circ$ (Jaffe et al.~1993).  On the other
hand, the narrow-line kinematic data from the central region suggests an inclination
angle of $69^\circ$ for the ten-parsec scale disk.  Both of these values are
consistent with the viewed angle of the radio jet, which is $\sim 63^\circ$ (Piner et
al.~2001).  We therefore adopt a value of $i = 64^\circ$ for the inclination angle of
the supersonic disk.

As stated in Section~\ref{sec:turbdom}, the radio-line absorption observations fix
the turbulent parameter at $f=0.045$.  To account for the radio shadowing near
$0.3$~pc, the electron number density should be around $30,000$~cm$^{-3}$.  Because
we expect hydrogen to be fully ionized in this region, we will choose $n_o =
30,000$~cm$^{-3}$ as the fiducial value for the scale-height-averaged number density of
the disk at a radius $r_o = 10^{18}$~cm from the supermassive black hole.  The 
remaining free parameters in the model are the ionizing luminosity $L_{\rm ion}$, and
the stress parameter $\alpha$, which determines the accretion rate and critical
radius via Equations~(\ref{mdot}) and (\ref{critical}).

\subsection{Methodology}

We use Cloudy to calculate the line emissivity as a function of radius for $r > r_c$,
including the additional heating (Eq.~\ref{eq:addheat}) and varying the density with
radius (Eq.~\ref{eq:density}) as predicted by our model. The line profiles are then
generated by integrating over the appropriate range in radius. We accomplish this
task by partitioning the disk into a large set of concentric tori, modeling the line
emission from each torus as a set of $N$ discreet clouds, with the understanding that
we approach the continuous limit with very large $N$. The emission line flux from the
population of $N$ clouds in torus $i$ as a function of wavelength is:
\begin{equation}
F_i(r, \lambda)\ \propto\ 2\pi\,r_i\,H(r_i)\,\Delta r_i\ \sum_{j=1}^N \ 
\epsilon({\bf r}_j,{\bf v}_j)\, \Lambda({\bf r}_j,{\bf v}_j,\lambda) \ ,
\label{eq:F_L}
\end{equation}
where the factor $2\pi\,r_i\,H(r_i)\,\Delta r_i$ is the volume of torus $i$,
$\epsilon({\bf r}_j, {\bf v}_j)$ is the line emissivity including relativistic
effects, and $\Lambda({\bf r}_j,{\bf v}_j,\lambda)$ is a convolution term relating a
cloud's location and velocity to the observed line wavelength. The composite line
flux from the entire disk is then obtained by summing the contributions from each
torus.

Taking the line-of-sight velocity to be $v_{\rm los}$, the effect of Doppler 
boosting is given by 
(\cite{MC97}):
\begin{equation}
\epsilon({\bf r},{\bf v})\ =\  \epsilon_0({\bf r})\ \left(\frac{\sqrt{1\ -\ 
v^2/c^2}}{1\ +\ v_{\rm los}^{\prime}/c}\right)^3 \ ,
\label{eq:epsilon}
\end{equation}
where $\epsilon_0({\bf r})$ is the rest-frame emissivity (generated by Cloudy) and
\begin{equation}
v_{\rm los}^{\prime}\ =\ v_{\rm los} + c \left( \sqrt{1 - \frac{r_S}{r}} - 1 \right)
\end{equation}
is the line-of-sight velocity including general-relativistic corrections.

The observed wavelength $\lambda$ of line emission from a given cloud is related to
rest-frame line wavelength $\lambda_0$ by
\begin{equation}
\lambda = \lambda_0\ \left(1 - \frac{r_S}{r}\right)^{-1/2} \left(\frac{\sqrt{1 - 
v^2/c^2}}{1- v_{\rm los}/c}\right) \ (1 + z)\;,
\label{eq:wavelength}
\end{equation}
where the correction terms on the right are the gravitational, Doppler, and
cosmological redshifts, respectively. This relation, appearing as a delta-function
$\Lambda({\bf r}_j,{\bf v}_j,\lambda)$ in Eq.~(\ref{eq:F_L}), is used to map each
cloud's emission into predetermined wavelength bins.

The axis of the disk is taken to lie along the $z$-axis. Cloud positions and
velocities are then determined via Monte Carlo sampling. Each cloud is given a random
azimuthal position $0 \leq \phi < 2\pi$ in the disk, and the velocity of each cloud
is assigned according to 
\begin{equation}
{\bf v}_j\ =\ v_k\, {\bf n}_\phi\ +\ {\bf v}_t \ ,
\end{equation}
where $v_k = (GM/r)^{1/2}$ is the Keplerian velocity, ${\bf n}_\phi$ is the azimuthal
direction vector (note that this depends on the value of $\phi$), and ${\bf v}_t =
\sqrt{f}\, v_k$ in a randomly-oriented direction. The line-of-sight velocity is then
equal to $v_{\rm los} = {\bf v}_j \cdot {\bf n}_{\rm los}$, where ${\bf n}_{\rm los}$
is the direction vector along the line of sight.

\subsection{Results of the Line Modeling}

\begin{figure}
\begin{center}
\resizebox{4in}{!}{\rotatebox{0}{\includegraphics{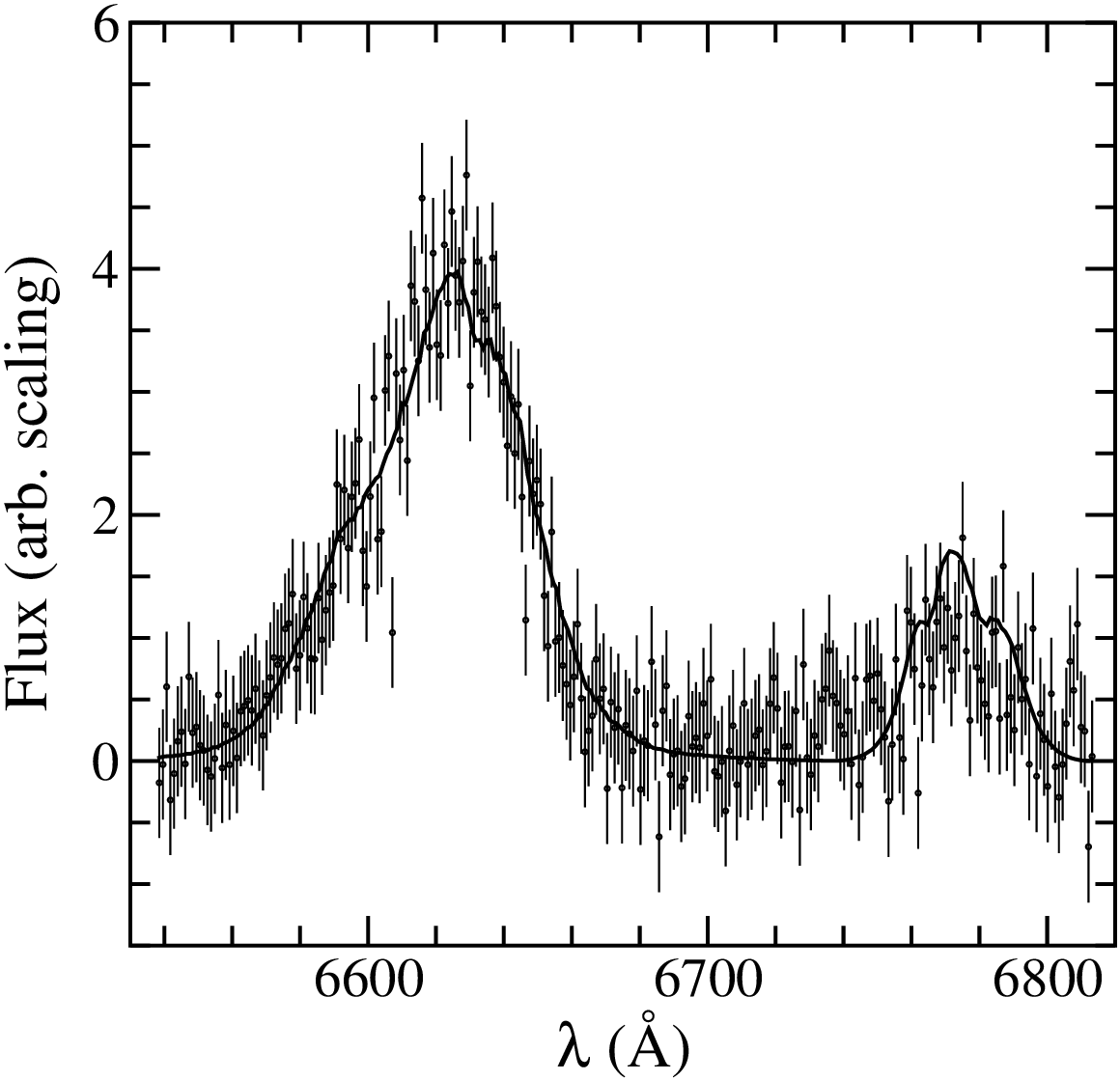}}}
\end{center}
\caption{
Observed (points with error bars) and best-fit modeled (solid curve) broad optical line 
spectrum of NGC
4261.  The best-fit free parameters are $L_{\rm ion} = 4 \times 10^{40}$~erg/s and 
$\alpha=0.4$,
resulting in $\chi_\nu^2=1.1$ ($\nu = 250$ degrees of freedom).  The other parameters are 
fixed in the
fit:  $n_o= 30000$~cm$^{-3}$, $f=0.045$, and $i = 64^{\circ}$ (see text for details).}
\label{fig:best_fit}
\end{figure}

We have modeled the continuum-subtracted H$\alpha$~$\lambda$6563 +
[\ion{N}{2}]~$\lambda\lambda$6548, 6584 and [\ion{S}{2}]~$\lambda\lambda$6717, 6731
broad emission-line spectrum of Ferrarese et al.~(1996). The parameters $L_{\rm
ion}$ and $\alpha$ were varied to determine the best $\chi^2$ fit to their nuclear
pixel data, corresponding to a FWHM spatial resolution of $0\farcs1 = 14$~pc.  
Figure~\ref{fig:best_fit} shows the resulting best fit ($\chi_\nu^2 = 1.1$; $\nu =
250$ degrees of freedom), having parameters $L_{\rm ion} = 4 \times 10^{40}$~erg/s
and $\alpha = 0.4$. The model nicely reproduces the broad wings and steep core of
the H$\alpha$ + [\ion{N}{2}] composite. There is an apparent excess in the modeled
[\ion{S}{2}] emission, which may be remedied by reducing the assumed abundance of
sulfur in the model (we used solar abundances throughout). Allowing for this
additional free parameter is beyond the scope of this paper, however, as we are more
interested in showing a consistency in the line profiles rather than their relative
strengths.

\begin{figure}
\begin{center}
\resizebox{6in}{!}{\rotatebox{270}{\includegraphics{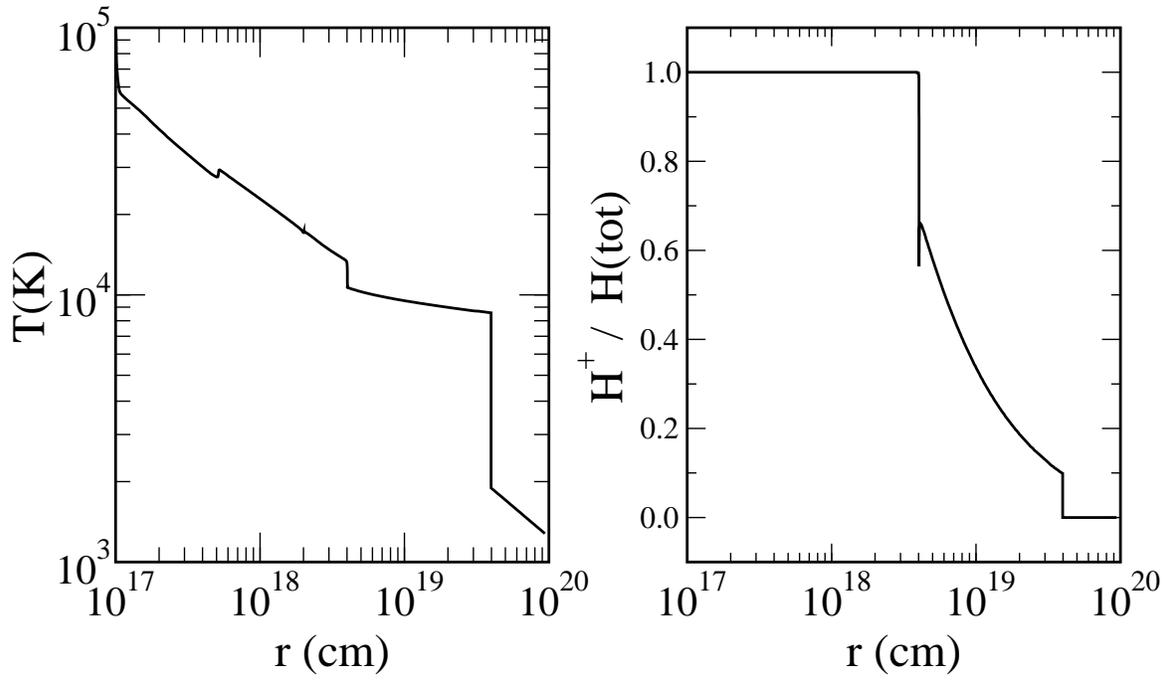}}}
\end{center}
\caption{The electron temperature (left) and ionization fraction (right) profiles of 
the disk.}
\label{fig:Temp_Ion}
\end{figure}

Figure~\ref{fig:Temp_Ion} shows the radial dependence of the gas temperature and the
ionization fraction of hydrogen in the best fit model. From this we see that most of
the hydrogen is ionized near $1$ pc from the center. So significant free-free
absorption is expected in the region. 

\begin{figure}
\begin{center}
\resizebox{4in}{!}{\rotatebox{270}{\includegraphics{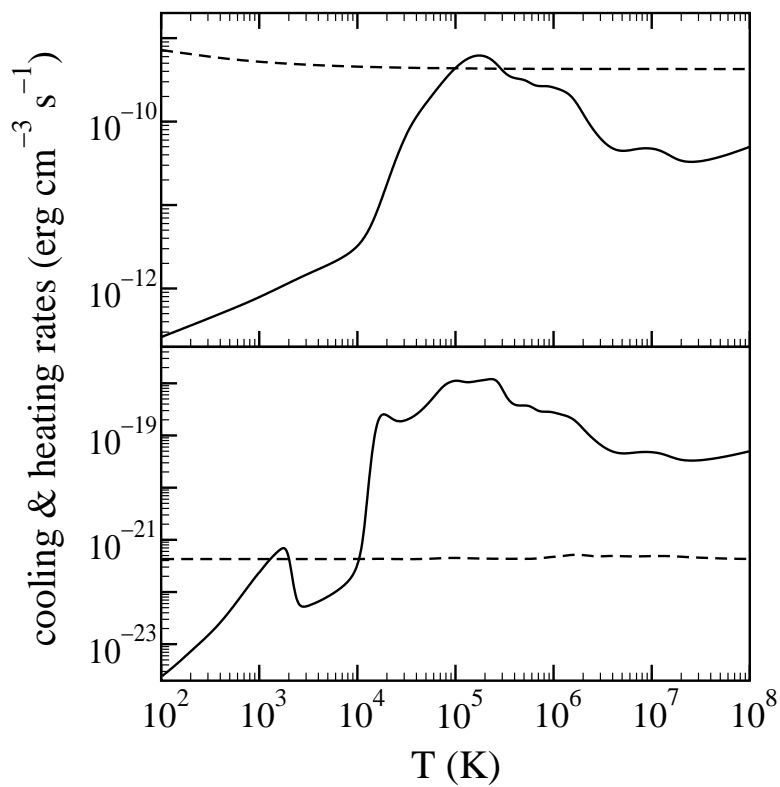}}}
\end{center}
\caption{Cooling (solid curves) and heating (dashed curves) rates in the disk gas at $r = 
10^{17}$~cm 
(top) and $r = 10^{20}$~cm (bottom). Note that the equilibrium temperature occurs where the 
heating and 
cooling rates balance. Stable equilibria occur only where the slope of the cooling function 
is 
positive. See text for details.
}
\label{fig:heat_cool}
\end{figure}

Thermal equilibrium occurs when the heating and cooling rates balance; stable
equilibria occur only where the slope of the cooling function is positive (e.g.,
Shore 1992). Figure~\ref{fig:heat_cool} illustrates the heating and cooling
functions for the gas at radii $r=10^{17}$~cm (top) and $r=10^{20}$~cm (bottom). The
heating rate includes both gravitational dissipation and radiative effects, and the
cooling rate must be determined including the dependency on the ionization state of
the gas. Cloudy calculates both of these rates in a self-consistent manner. At
$r=10^{17}$~cm, it is clear that only one stable solution exists, $T \simeq 10^5$~K
(compare with Fig.~\ref{fig:Temp_Ion}), so there is no ambiguity in determining the
equilibrium temperature. At large radii, however, it is apparent that two valid
physical solutions exist. Because our picture has the gas being heated from a low
temperature as it accretes, we assign the smaller value in all such cases of
uncertainty. As shown in figure~\ref{fig:Temp_Ion}, the gas temperature jumps from
$\sim 2\times 10^3$~K to $\sim 10^4$~K near $4\times 10^{19}$~cm when the gas moves
from the lower temperature stable branch to the higher temperature one. This is
interesting because neutral atomic hydrogen is concentrated at that radius and this
may therefore explain the detected neutral hydrogen radio line absorption.

\begin{figure}
\begin{center}
\resizebox{4in}{!}{\rotatebox{270}{\includegraphics{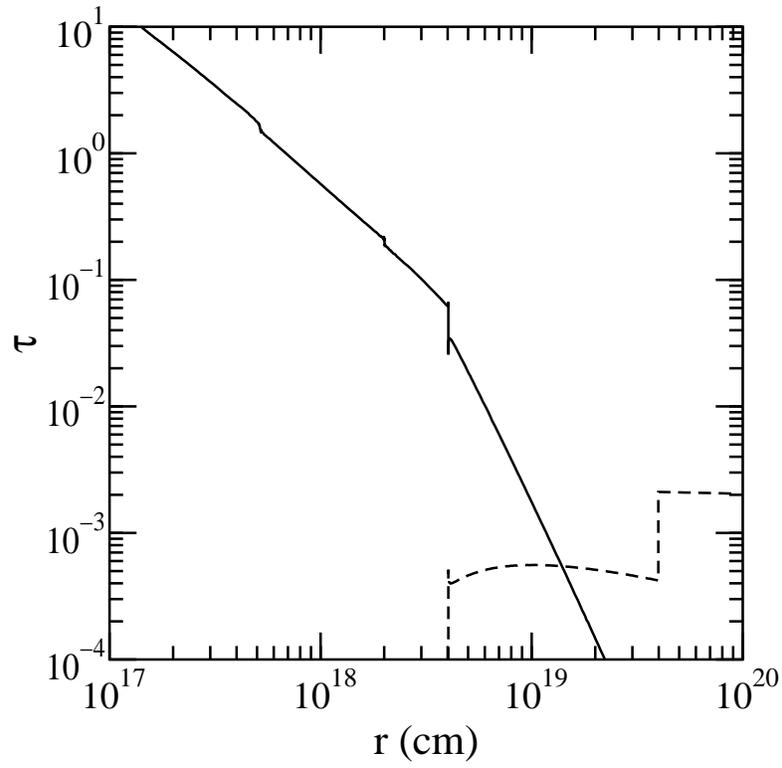}}}
\end{center}
\caption{The free-free optical depth $\tau_\nu^{\rm ff}$ at $5$~GHz (solid curve) and the 
line-center 
optical depth $\tau_0^{\rm 21\, cm}$ of \ion{H}{1}~$\lambda 21$~cm (dashed curve), 
calculated for the supersonic disk viewed face-on. 
}
\label{fig:opt_depth}
\end{figure}

We have shown that our model can reproduce the observed broad-line spectrum; we now
turn to the question concerning the observed absorption features. The free-free absorption
optical depth at frequency $\nu$ is given by (Walker et al.~2000):
\begin{equation}
\tau_\nu^{\rm ff}\ =\ (9.8\times 10^{-3})\ n_e^2\ l\, T^{-1.5}\, \nu^{-2}\ \left[\, 
17.7\ +\ \ln{(T^{1.5}\, \nu^{-1})}\, \right] \ ,
\end{equation}
where all the quantities are given in cgs units.  
The line-center \ion{H}{1}~$\lambda21$~cm optical depth, corrected for stimulated 
emission, is given by 
(\cite{Rohlfs00}):
\begin{equation}
\tau_0^{21\, {\rm cm}}\ =\ \left(1.45 \times 10^{-15}\ {\rm cm^{2}\ Hz\ K} \right)\ 
\frac{N_H}{\Delta \nu_D\, T} \ ,
\label{eq:tau21}
\end{equation}
where $N_H$ is the neutral hydrogen column depth and
\begin{equation}
\Delta \nu_D\ =\ \frac{1}{\lambda}\ \sqrt{\frac{2\, k T}{m_H}\ +\ v_{t}^2}
\end{equation}
is the Doppler line width (including both turbulent and thermal broadening). Note
that for $r > r_c$, the turbulent velocity of the gas is always much greater than
the thermal velocity. Figure~\ref{fig:opt_depth} shows the radial dependence of both
of these optical depths for our best-fit model, under the assumption that the line
of sight is perpendicular to the disk. The free-free absorption is strong enough to
account for the observed radio shadowing. However, the neutral hydrogen absorption
seems deficient. This is not surprising given that we have neglected the density
contrast of the supersonic disk in the calculation. Observations of other
radio galaxies have revealed that the gas is clumpy at the same spatial scale (e.g.,
Harms et al. 1994; Haschick, Baan \& Peng 1994; Kartje, K\"{o}nigal \& Elitzur
1999).  Introducing density contrast will not only change the temperature profile of
the gas, but also increase the neutral hydrogen column density. So we would expect
significant enhancement of radio line absorption in a detailed modeling that
incorporates these additional effects.

\begin{figure}
\begin{center}
\resizebox{4in}{!}{\rotatebox{0}{\includegraphics{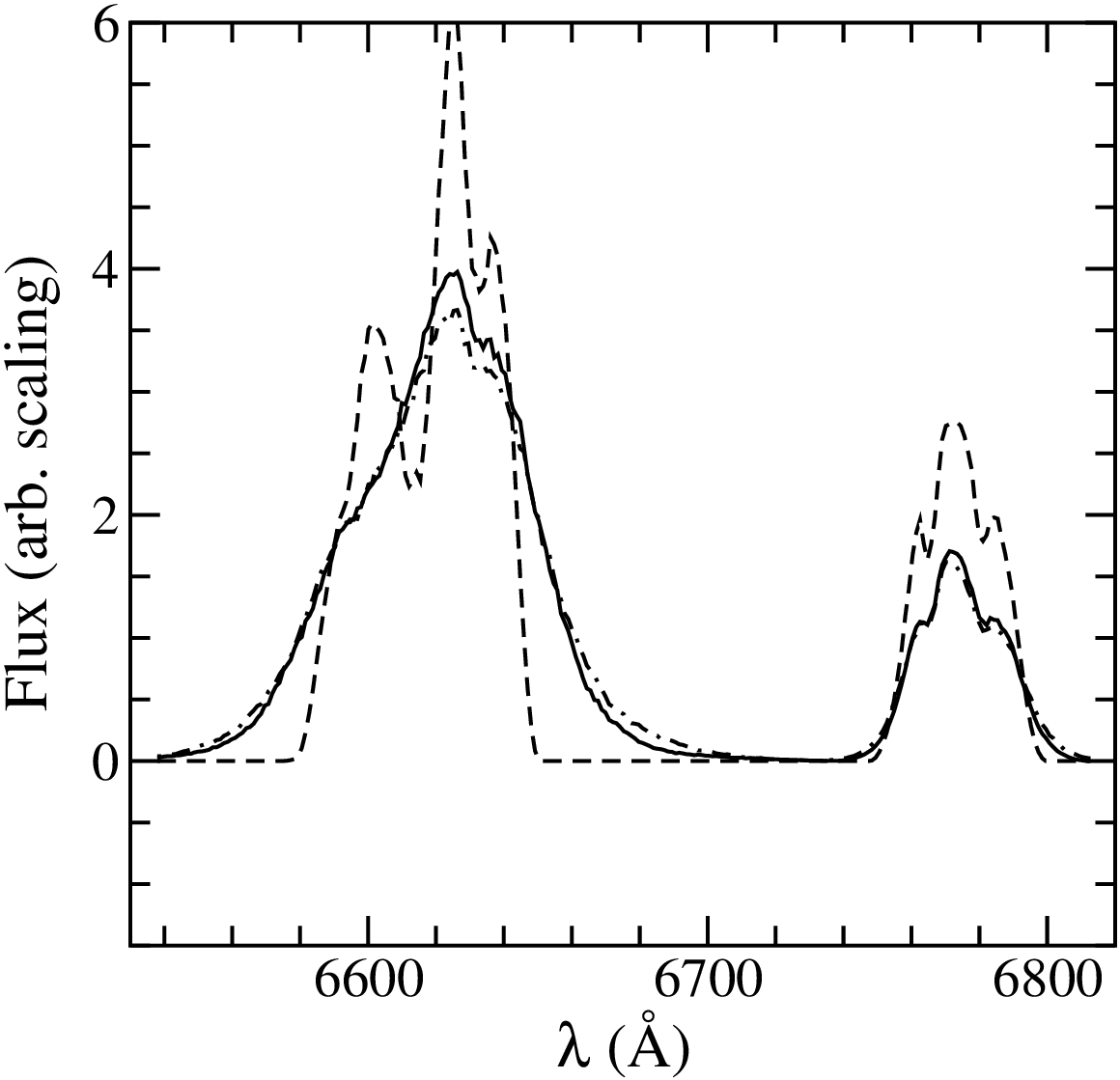}}}
\end{center}
\caption{The dependence of the broad-line spectrum on the ionizing luminosity 
$L_{\rm ion}$.  Shown in the figure are the modeled spectra for $L_{\rm ion} = 4 \times 
10^{39}$~erg/s (dot-dashed curve), $L_{\rm ion} = 4 \times 10^{40}$~erg/s (solid 
curve), and $L_{\rm ion} = 4 \times 10^{41}$~erg/s (dashed curve).
The other parameters are fixed at:  $\alpha = 0.4$, $n_o= 30000$~cm$^{-3}$, $f=0.045$, and 
$i = 
64^{\circ}$.
}
\label{fig:vary_L}
\end{figure}

Figure~\ref{fig:vary_L} illustrates how the ionizing flux modifies the line
spectrum.  With an increase in $L_{\rm ion}$, the heating and ionization parameters
of the gas increase as well. This has the effect of moving the emission zone to larger
radii, particularly for the low ionization [\ion{N}{2}] and [\ion{S}{2}] species. As
a result, the line profiles become narrower for increasing $L_{\rm ion}$. A
reduction of $L_{\rm ion}$ from the best fit value does not have such a drastic
impact, however. This is because the inner radius reaches an asymptotic limit in the
low luminosity case, determined solely by balancing the gravitational heating and
line cooling rates via Equation~(\ref{critical}). In this limit, the additional heating
due to the radiation field is negligible. It is interesting to note that our best fit
lies on the transition to this limit.

\begin{figure}
\begin{center}
\resizebox{4in}{!}{\rotatebox{0}{\includegraphics{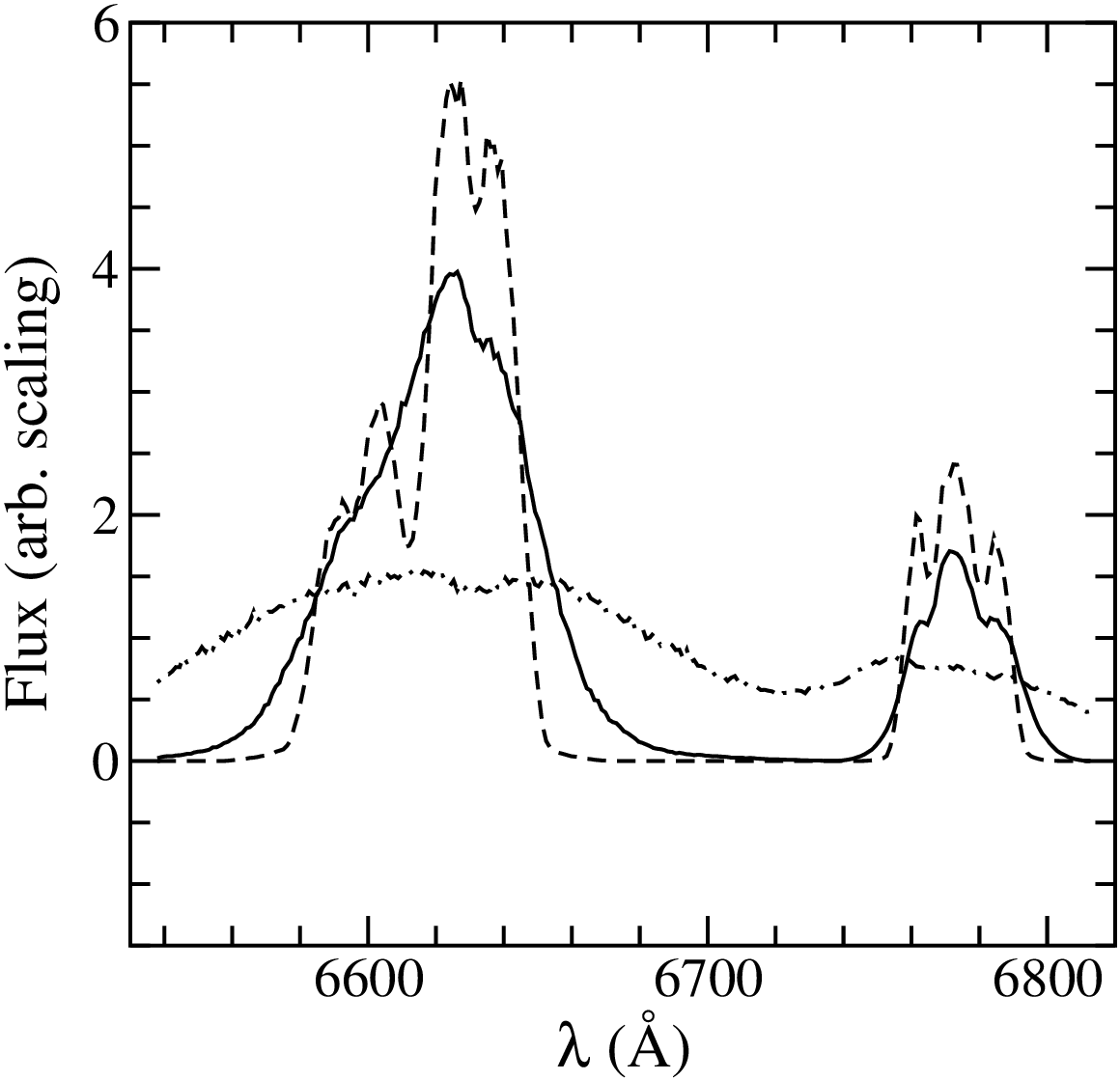}}}
\end{center}
\caption{The dependence of the broad-line spectrum on the stress parameter
$\alpha$.  Shown in the figure are the modeled spectra for $\alpha=0.04$ (dot-dashed 
curve), $\alpha=0.4$ (solid curve), and $\alpha=1$ (dashed curve).
The other parameters are fixed at:  $L_{\rm ion} = 4 \times 10^{40}$~erg/s, $n_o= 
30000$~cm$^{-3}$, 
$f=0.045$, and $i = 64^{\circ}$.
}
\label{fig:vary_a}
\end{figure}

The dependence of the line shape on the stress parameter $\alpha$ is demonstrated
in Fig.~\ref{fig:vary_a}. A larger $\alpha$ increases the accretion rate (for a
fixed number density at $r_o$), causing
additional heating in the gas (via Eq.~\ref{eq:addheat}). This moves the emission
zone to larger radii, resulting in narrower line profiles. Decreasing $\alpha$ has
the opposite effect.

\begin{figure}
\begin{center}
\resizebox{4in}{!}{\rotatebox{0}{\includegraphics{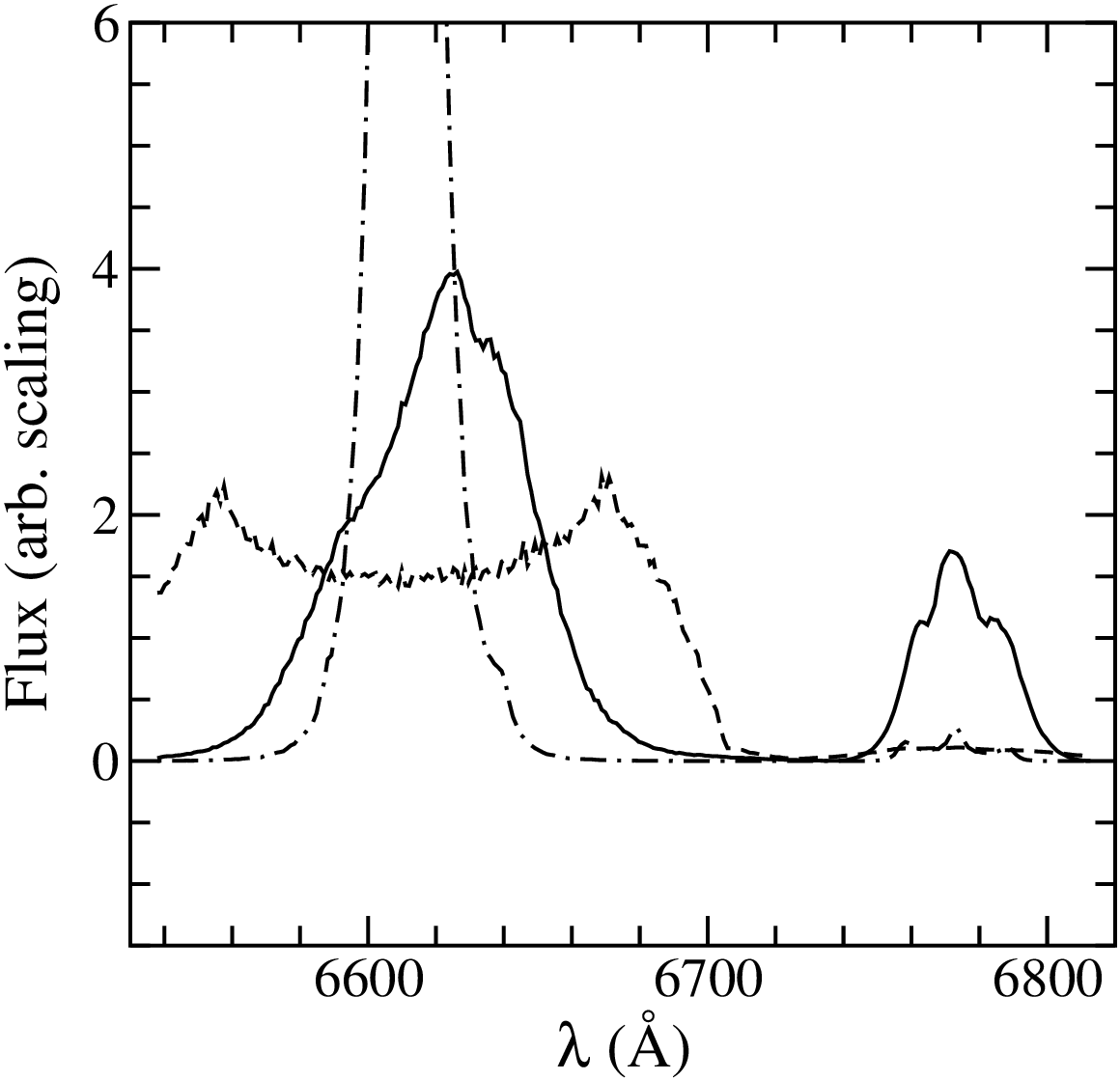}}}
\end{center}
\caption{The dependence of the broad-line spectrum on the number density $n_o$ of the disk 
at a distance of $10^{18}$ cm from the black hole.  Shown in the figure are the modeled 
spectra for $n_o=3000$~cm$^{-3}$ (dot-dashed curve), $n_o=30,000$~cm$^{-3}$ (solid 
curve), and $n_o=300,000$~cm$^{-3}$ (dashed curve).
The other parameters are fixed at:  $L_{\rm ion} = 4 \times 10^{40}$~erg/s, $\alpha = 0.4$, 
$f=0.045$, 
and $i=64^{\circ}$.
}
\label{fig:vary_n}
\end{figure}

We have used the observed radio free-free absorption to fix our value of $n_o$ in
the best fit model. Figure~\ref{fig:vary_n} illustrates the effect of varying $n_o$.
The density of the gas is important in many aspects, including a determination of
the ionization parameter, as well as the gravitational heating and line cooling
rates. It also has a strong impact on the fluxes of the [SII] and [NII] forbidden
lines, which are suppressed relative to H$\alpha$ for densities greater
than the critical density, at which the radiative and collisional transition rates
balance.

\begin{figure}
\begin{center}
\resizebox{4in}{!}{\rotatebox{0}{\includegraphics{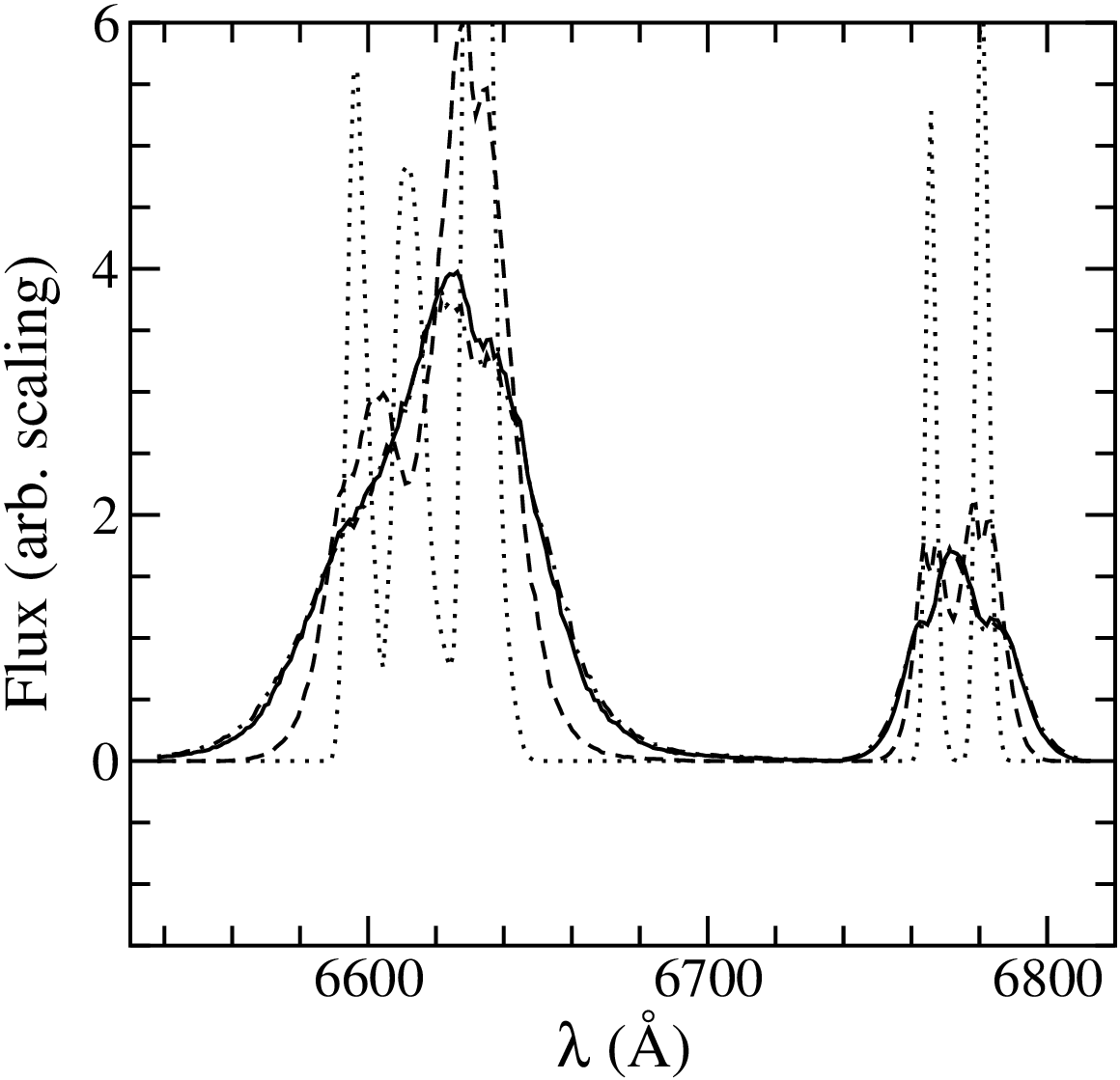}}}
\end{center}
\caption{The dependence of the broad-line spectrum on the inclination angle of the disk 
$i$.  
Shown in the figure are the modeled spectra for $i=0^{\circ}$ (dotted curve), 
$i=30^{\circ}$ (dashed curve), $i=64^{\circ}$ (solid curve), and $i=75^{\circ}$ (dot-dashed 
curve).
The other parameters are fixed at:  $L_{\rm ion} = 4 \times 10^{40}$~erg/s, $\alpha = 0.4$, 
$n_o= 
30000$~cm$^{-3}$, and $f=0.045$.
}
\label{fig:vary_i}
\end{figure}

Because the line emission broadening is due mainly to the Keplerian motion of the
gas, the line width is proportional to $v_k \sin{i}$, where $v_k$ is the Keplerian
velocity in the disk. The projection effect, characterized by the
inclination angle $i$, is demonstrated in Fig.~\ref{fig:vary_i}. As expected, model
spectra associated with small inclination angles have very narrow line profiles. Note, 
however,
that even when $i=0$ the lines have non-zero width. This is the effect of broadening
due to the turbulent velocity $v_t = \sqrt{f} v_k$.

\begin{figure}
\begin{center}
\resizebox{4in}{!}{\rotatebox{0}{\includegraphics{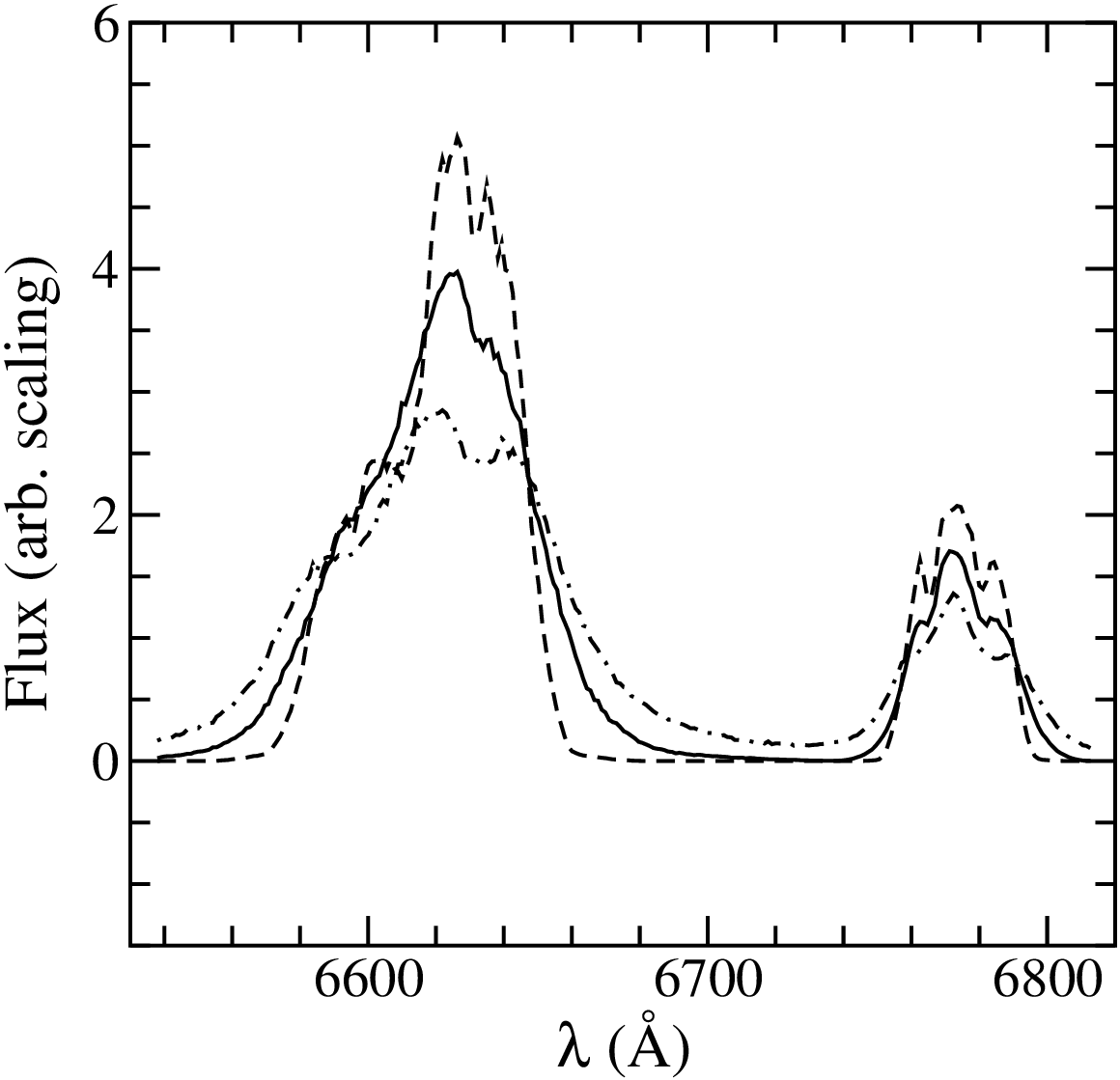}}}
\end{center}
\caption{The dependence of the broad-line spectrum on the turbulence parameter $f$.  
Shown in the figure are the modeled spectra for $f=0.011$ (dot-dashed curve), $f=0.045$ 
(solid curve), and $f=0.180$ (dashed curve).
The other parameters are fixed at:  $L_{\rm ion} = 4 \times 10^{40}$~erg/s, $\alpha = 0.4$, 
$n_o= 
30000$~cm$^{-3}$, and $i=64^{\circ}$.
}
\label{fig:vary_f}
\end{figure}

Finally, the turbulent parameter $f$ has a similar effect on the accretion rate
as the stress parameter $\alpha$. However, by increasing $f$, we not only
shift the broad emission line region toward large radii, but also increase the total
line flux, as the disk becomes thicker (thereby increasing the size of the emission
region).  Figure~\ref{fig:vary_f} shows how the line spectrum changes with $f$.

\section{Discussion}
\label{discussion}

The accretion flow within $r_t$ is not well constrained.  The fact that the inferred
accretion rate from this supersonic disk model gives a radiative efficiency of $\sim
2\times 10^{-4}$ suggests that advection may dominate in the inner region (Narayan  
\& Yi, 1994). However, if we associate the observed X-ray emission with such
an advection dominated accretion flow (ADAF) at small radii, we find that the observed 
X-ray spectrum with a spectral index of $\sim 0.4$ (Rinn \& Sambruna 2001; Matsumoto et al
2001) is somewhat softer than that of the model prediction (Narayan et al. 1998). An
ADAF with strong winds may be able to produce a softer X-ray spectrum to account for 
the high energy emission from the AGN in NGC~4261 (Quataert \& Narayan 1999). And 
this outflow may be associated with the jet formation process in the core.

It is also interesting to note that NGC~6251 is quite similar to NGC~4261 in that both
of them have a hundred-pc scale dusty disk, large scale symmetric radio jets and an
inverted radio spectrum at the core (Melia, Liu \& Fatuzzo 2002).  Compared with NGC
4261, NGC~6251 is about 3 times further away from us, its X-ray luminosity is $30$
times stronger, and its broad H$\alpha$ line flux is about 100 times bigger. HST
observations suggest that there is a supermassive black hole of $6\times 10^8\msun$,
which is very close to the black hole mass in NGC~4261. If a similar accretion flow
exists in NGC~6251, Equation (\ref{line}) suggests that the electron number density
must increase by a factor of $\sim 10$ to produce the strong H$\alpha$ emission.
Combining this with the fact that the disk has an inclination angle of $\sim 32^\circ$ 
(Ferrarese \& Ford 1999) and that the AGN is more active, we would expect that the
observed broad emission line may also be fitted with our model. We also would expect
strong radio line absorption in the direction of the counterjet with an H~{\small
I} column density of $\sim 10^{23}$ cm$^{-2}$.

In an earlier paper we introduced a hot expanding gas model for the multi-wavelength 
spectrum of the AGN in NGC~6251 (Melia et al. 2002). According to that picture, a 
standard $\alpha$-disk with an accretion rate of $\sim 4\times 10^{22}\,$g~s$^{-1}$ 
exists near the black hole. How such a fossil disk can form after the 
supersonic disk terminates is still an open question. However, the 
implied strong evaporation of the accretion flow below $r_t$ suggests another origin 
for the hot unbound gas proposed in that model. Further investigation is clearly
warranted, and an application of the supersonic disk model to other radio 
galaxies may also prove to be invaluable in developing a unified theory for 
supermassive black hole accretion.

{\bf Acknowledgments} 

This research was partially supported by NASA under grants NAG5-8239 and NAG5-9205,
and has made use of NASA's Astrophysics Data System Abstract Service.  FM is very
grateful to the University of Melbourne for its support (through a Miegunyah
Fellowship).  MJF would like to thank Gary Ferland for his helpful suggestions.

{}

\clearpage

\end{document}